# The Opium for the Poor Is – Opium.
# Medicare Providers in States with Low Income and Republican Control Prescribe High Levels of Opiates

Eugen Tarnow, Ph.D.

Avalon Business Systems, Inc.

AvalonAnalytics.com

Avalon Working Paper 071316

19-03 Maple Avenue, Fair Lawn, NJ 07410

+1 646 825-9080

etarnow@avabiz.com

## Abstract

The majority of Medicare opioid prescriptions originate with family practice and internal medicine providers. I show that the average number of Medicare opium prescriptions by these providers vary strongly by state and that 54% of the variance is accounted for by the state median household income. I also show that there is a very similar relationship in opioid claims per capita and per Medicare recipient. In all cases Alabama is the state with the most claims and Hawaii is the state with the least claims. I also show that there is a significant difference depending upon party in control: states with Republican governors, majorities in upper or lower houses have higher opioid claims. *Keywords*: opioid prescriptions, family practice, internal medicine, average income, by state, Medicare

# The Opium for the Poor Is – Opium. Medicare Providers in States with Low Income and Republican Control Prescribe High Levels of Opiates

The current opiate epidemic has many reasons: pharmaceutical marketing, naïve and/or unscrupulous doctors, availability of cheap and cleverly sold Mexican heroin and inappropriate policy changes to encourage the idea that pain was something curable with drugs that were not addictive (Quinones, 2015).

In this paper I investigate how opioid claims vary by state.

## Definitions & Data Sources

I analyze the Medicare provider datasets for the presence of opioid prescription claims (Medicare_Part_D_Opioid_Prescriber_Summary_File.csv can be downloaded from data.cms.gov). Different types of providers will have different reasons to prescribe opioids. To simplify the analysis, I limit it to family doctors and internal medicine doctors, they are responsible for most opioid prescriptions (see Table 1). To avoid statistical noise, I further limit the investigations to those doctors with at least 200 Medicare claims per year.

As a measure of state income I use the median household income from 2014, as downloaded from Wikipedia (downloaded 8/24/16 from https://en.wikipedia.org/wiki/List_of_U.S._states_by_income).

The number of Medicare recipients per state was downloaded from https://www.cms.gov/Newsroom/MediaReleaseDatabase/Press-releases/2015-Press-releases-items/2015-07-28.html on 8/25/16.

**Results & Discussion**

The state average of Medicare opiate claims per provider varies between 528 claims (District of Columbia) all the way up to 15,000 claims (Alabama).  The state average of Medicare opiate claims per provider as a function of the state median household income is displayed in Fig. 1.  The relationship accounts for 54% of the variance in prescriber rates. In other words, if you live in a poor state, it appears that you are much more likely to be prescribed opiates by your Medicare physician than if you live in a rich state.

To check the result, I also calculated the per capita number of opioid claims in each state and obtained the very similar result in Fig. 2.

In both cases the functional form is a power law ~(Median household income)$^{-n}$ where n varies from 2.4 to 2.9.

It is clear that Medicare is not supervising the opiate claims.  This is not surprising since I have found before that Medicare supervision fails when it comes to volume discounts for heart attacks, quality for higher heart attack payments, and pricing (overpays its providers; see blog at analyticsguru.us) and Medicare fraud is an all too common topic in the press.  Indeed, Medicare's attempt to supervise seems to be to release datasets and wait for others to do the supervision.

It is not clear what the reason behind the relationship between opioid claims and median household income. Is it that people in poor states tend to display their suffering higher up on the pain scale, or they benefit from or feel the need for more opioids?  Is it that doctors in poor states need to create more patients for their practices?  Is it that people in richer states get the same amount of opioids but can afford to get it from sources other than Medicare?

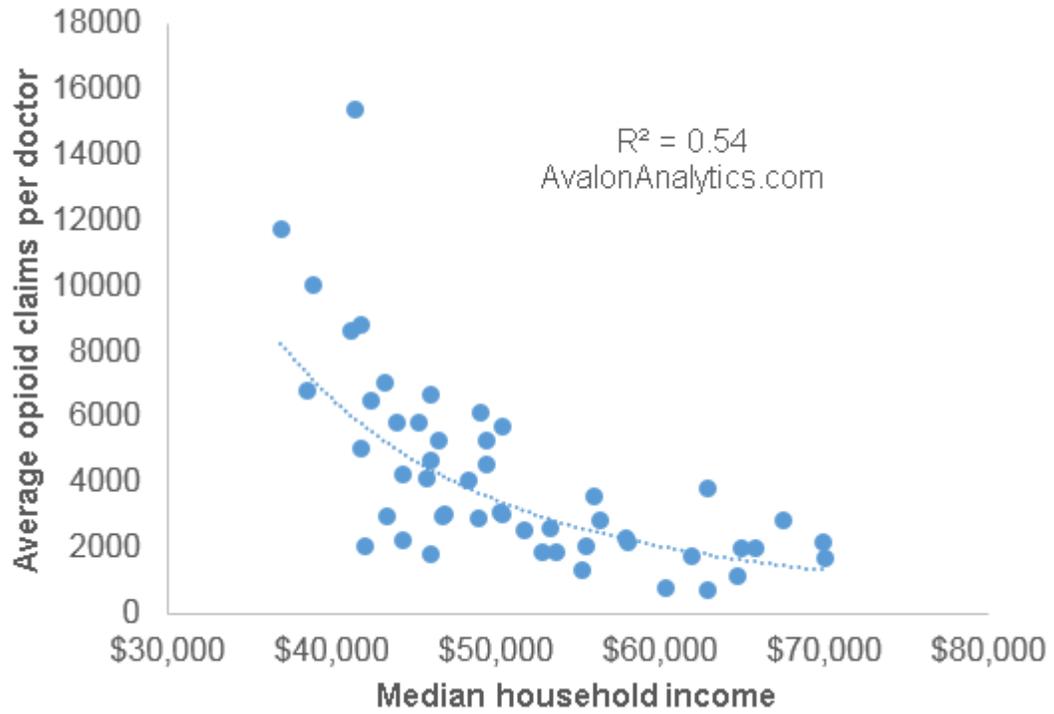

*Figure 1.  State average number of opioid claims by provider as a function of the median household income.  The highest prescribing provider state is Alabama and the lowest is Hawaii.*

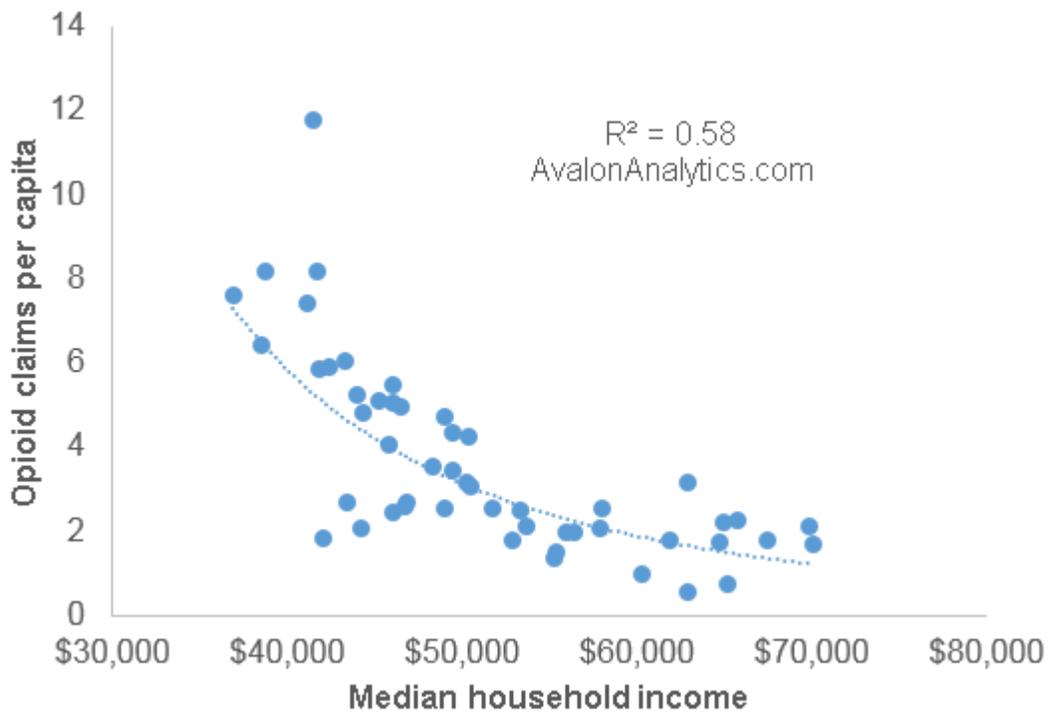

*Figure 2. Medicare Opioid claims per capita as a function of the median household income. The state with the highest number of opioid claims per capita is Alabama and the lowest number is Hawaii.*

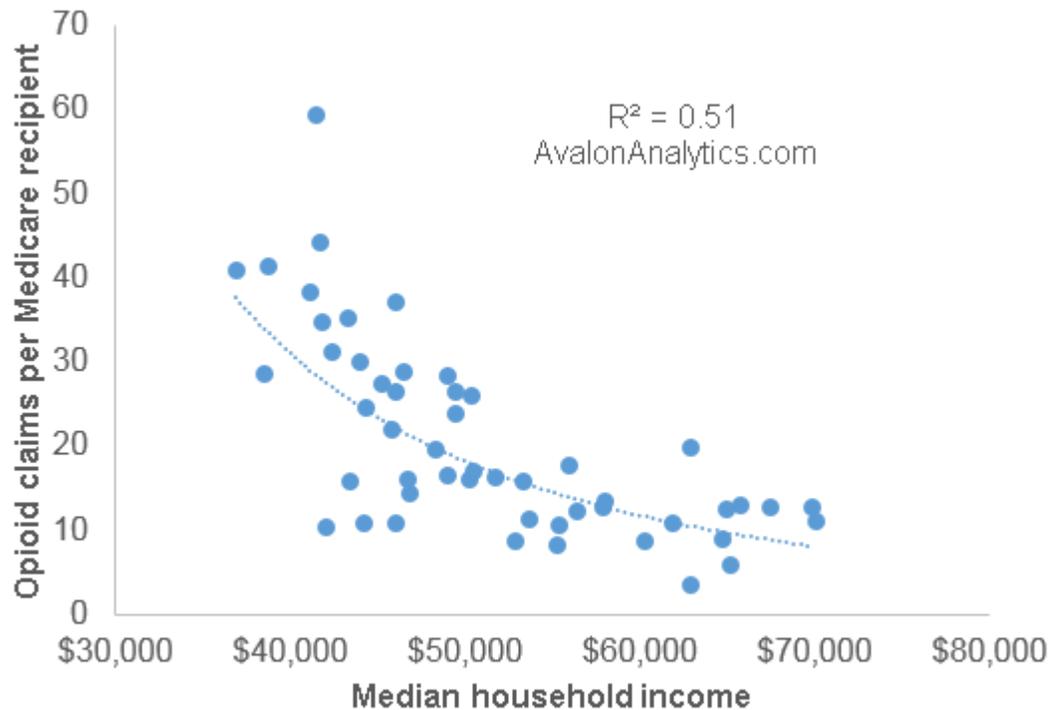

*Figure 3. Medicare Opioid claims per Medicare recipient as a function of the median household income. The state with the highest number of opioid claims per Medicare recipient is Alabama and the lowest number is Hawaii.*

A possible reason for the relationship of poor states and high opioid prescriptions is that state legislation may impact prescription behavior and poor states may be more susceptible to the largess of pharmaceutical lobbying than rich states. If that is the case, there might be a difference in prescriptive behavior also with party in power.

It turns out that states with Republican legislatures have providers that write more Medicare opioid prescriptions. This is shown in Fig. 4. In each of the panels is displayed the percent Medicare opioid claims out of all Medicare claims as a function of median household income. The red dots correspond to Republican control and the blue dots to Democratic control. The lines are the least square fits to the dots. The top panel shows that states with Republican governors have more Medicare opioid claims than states with Democratic governors. The middle panel shows that is even more so for Republican upper houses and the bottom panel shows that it is even more so for Republican lower houses (I used the Wikipedia definition of lower and upper houses).

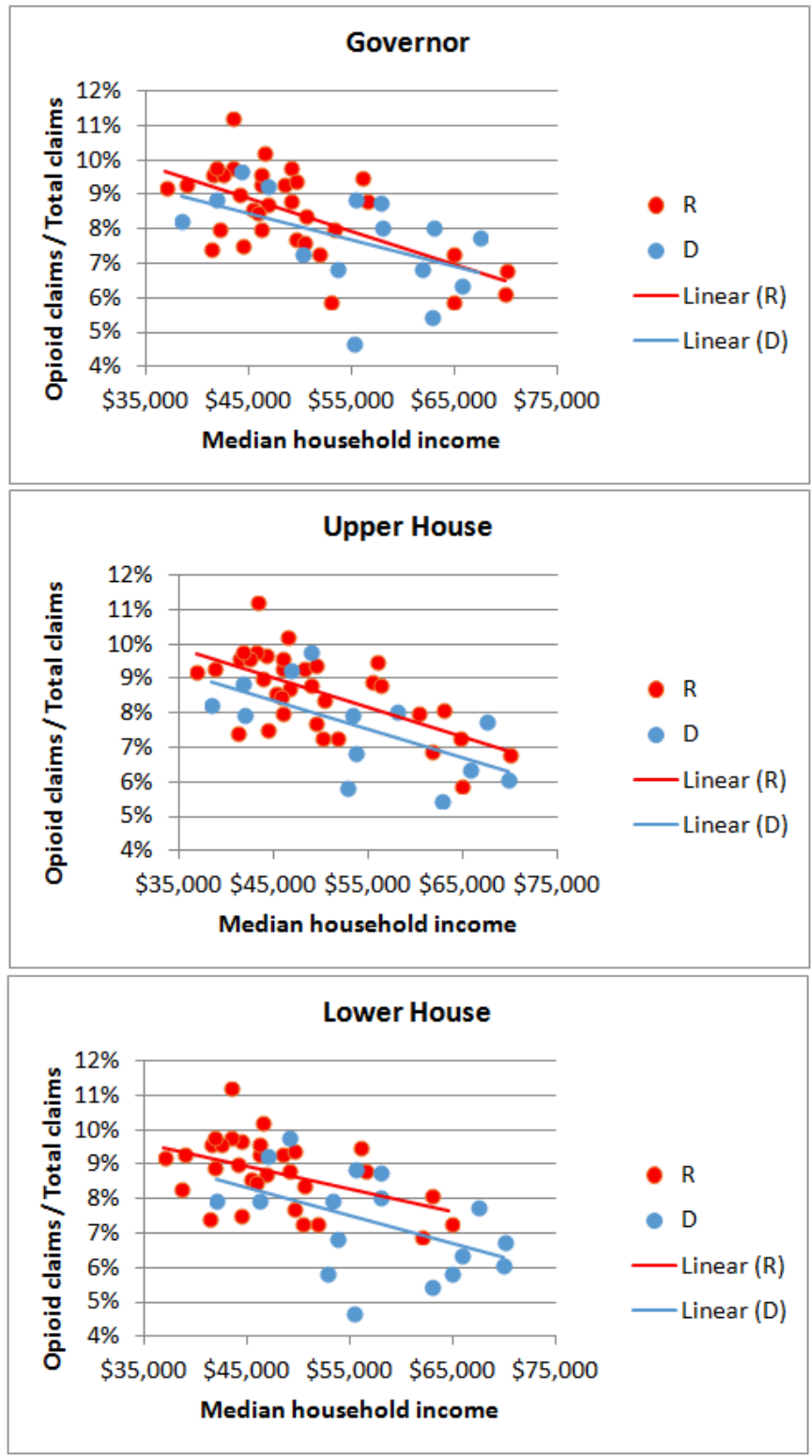

*Fig. 4. Opioid claims as a percentage of all Medicare claims as a function of median household income and party controlling the state government.*

This result is not obvious.  In Fig. 5 are the data displayed for the murder rate.  The murder rate is a lot less sensitive to the median household income and the variation with party is not as consistent.

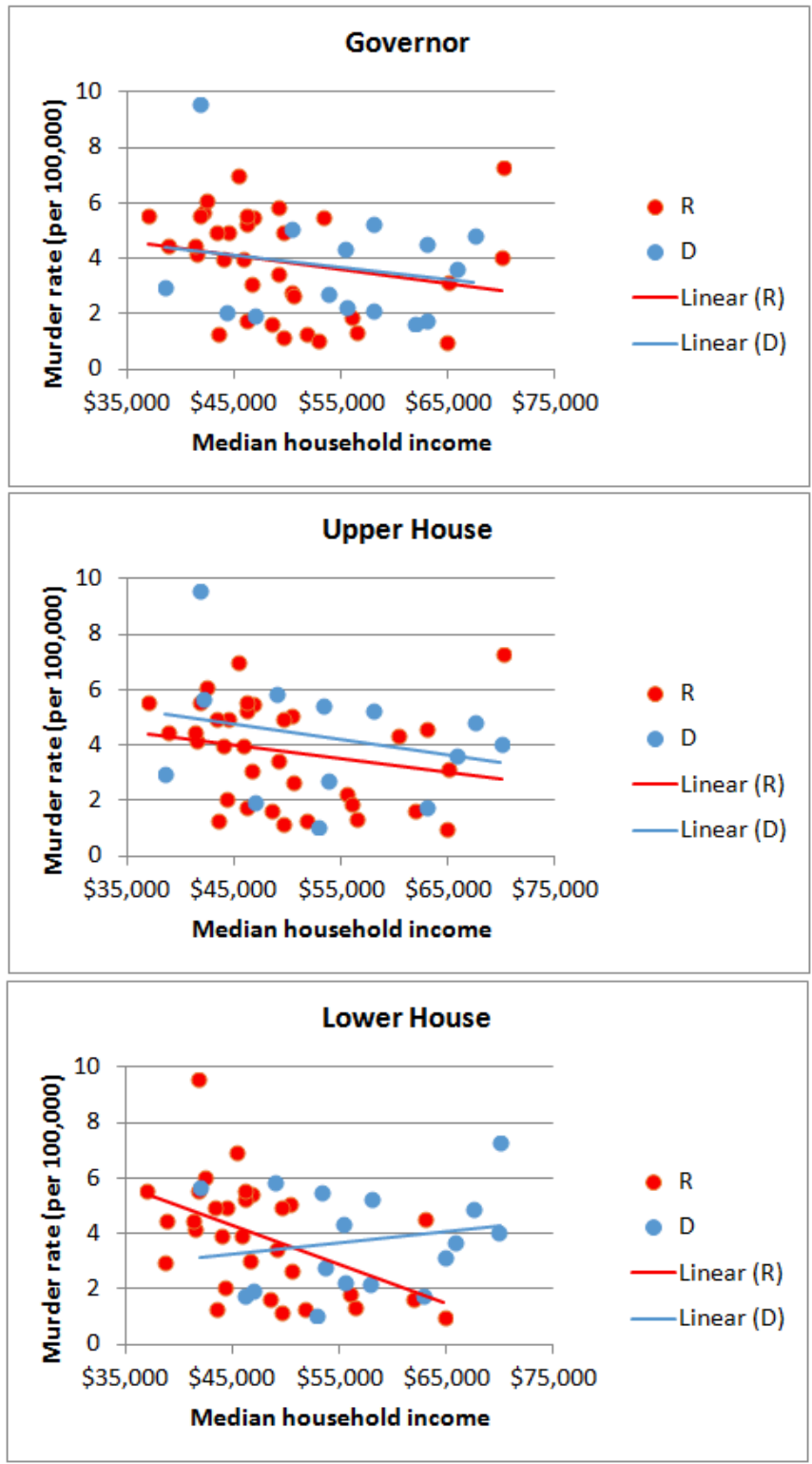

*Fig. 5. Murder rate as a function of median household income and party in control of the state government.*

OPIUM FOR THE POOR IS – OPIUM													11A one-way ANOVA showed that there was a significant difference between Democrat and Republican Lower Houses for opioids (F=16.7, p<0.001) but not for the murder rate (p=0.8), Upper Houses for opioids (F=4.6, p=0.015) but not for the murder rate (p=0.67), Governor for opioids (F=4.23, p=0.045) but not for the murder rate (p=0.58).

OPIUM FOR THE POOR IS – OPIUM	12References

Quinones, S. (2015). Dreamland: The true tale of America's opiate epidemic. *Health Affairs*, *34*, 9..

.

Tables

Table 1

*Medicare opioid claims by provider specialty*

| Specialty Description | Opioid Claims | Percent |
|---|---:|---:|
| Family Practice | 20234606 | 28.2% |
| Internal Medicine | 17548868 | 24.4% |
| Nurse Practitioner | 4952103 | 6.9% |
| Physician Assistant | 3725719 | 5.2% |
| Orthopedic Surgery | 3234667 | 4.5% |
| Physical Medicine and Rehabilitation | 2585335 | 3.6% |
| Anesthesiology | 2292388 | 3.2% |
| Interventional Pain Management | 2271295 | 3.2% |
| Emergency Medicine | 2099521 | 2.9% |
| General Practice | 1537293 | 2.1% |
| Rheumatology | 1394614 | 1.9% |
| Pain Management | 1353319 | 1.9% |
| Neurology | 984190 | 1.4% |
| General Surgery | 870633 | 1.2% |
| Dentist | 846999 | 1.2% |